\documentclass[9pt,twocolumn,twoside]{osajnl}

\journal{josaa} 

\setboolean{shortarticle}{true} 

\title{Theoretical study  of laser intensity noise effect on CW-STED microscopy}

\author[1]{Alejandro Mendoza-Coto}
\author[2]{Danay Manzo Jaime}
\author[3]{Ariel Francis P\'erez Mellor}
\author[4,*]{Iv\'an Coto Hern\'andez}

\affil[1]{Departamento de Física, Universidade Federal de Santa Catarina, 88040-900 Florianópolis, Santa Catarina, Brazil}

\affil[2]{Department of Mechanical Engineering, Universidade Federal de Santa Catarina, Florianópolis, Brazil}

\affil[3]{Department of Physical Chemistry, University of Geneva, Switzerland}

\affil[4]{Surgical Photonics and Engineering Laboratory, Mass Eye and Ear and Harvard Medical School, 243 Charles St., Boston, MA, USA 02114}

\affil[*]{Corresponding author: ivancotohernandez@meei.harvard.edu}

\ociscodes{(140.3460) Lasers; (180.2520)   Fluorescence microscopy.}

\begin{abstract}
Spatial resolution of stimulated emission depletion (STED) microscopy varies with sample labeling techniques and microscope components, e.g.,
lasers, lenses, and photo-detectors. Fluctuations in the intensity of the depletion laser decrease achievable resolution in STED microscopy; the stronger the fluctuations, the higher the average intensity needed to achieve a given resolution. This phenomenon is encountered in every STED measurement. However, a theoretical framework that evaluates the effect of intensity fluctuations on spatial resolution is lacking. 
This article presents an analytical formulation based on a stochastic model that characterizes the impact of the laser fluctuations and correlation time on the depletion efficiency in the continuous wave (CW) STED microscopy. We compared analytical results with simulations using a wide range of intensity noise conditions and found a high degree of agreement. The stochastic model used considers a colored noise distribution for the laser intensity fluctuations. Simple analytical expressions were obtained in the limit of small and large fluctuations correlation time. These expressions fitted very well the available experimental data. Finally, this work offers a starting point to model other laser noise effects in various microscopy implementations.
\end{abstract}

\setboolean{displaycopyright}{true}

\begin{document}

\maketitle

Stimulated emission depletion (STED) microscopy 
\cite{Hell1994,10.1038/nmeth.4593} is a revolutionary imaging technique that provides a means of bypassing the diffraction limit of light, allowing biological samples to be observed in great detail.  \cite{vicidomini2014gated}. Its use has been widely explored, giving rise to multiple implementations and a growing number of biomedical publications. To improve the optical resolution of confocal microscopy, STED employs a second laser to deplete the fluorescence in the outer region of the diffraction-limited spot produced by the excitation laser. A streamlined version of STED microscopy can be implemented with continuous-wave (CW) lasers, reducing tedious laser pulse preparation \cite{Willig2007}. In addition, the progress of laser technology in the last two decades has allowed the implementation of CW-STED microscopy with inexpensive and compact lasers, such as visible-fiber lasers \cite{Eggeling2012, coto2014new}, diode-pumped-solid-state lasers (DPSLs) \cite{mueller2012cw} and optical pumped-semiconductor lasers (OPSLs) \cite{vicidomini2014gated}.  

Quantum and technical noises from laser sources generate intensity fluctuations  \cite{paschotta2009noise1} that can affect image contrast. Outstanding results in STED microscopy have been obtained with low-noise DPSS and OPSL lasers \cite{Eggeling2012, vicidomini2014gated}: which have a root mean square noise less than 0.1 \%. Coto Hernández, I. et al. \cite{coto2014influence} have experimentally shown that the use of a low-noise STED laser reduces the intensity needed to achieve super-resolution.  
Laser noise can be characterized by its frequency-dependent power spectral density, often measured with a photodiode and electronic spectrum analyzer \cite{coto2014influence}. Furthermore, it can be analyzed through its intensity fluctuation or correlation time. White noise, uncorrelated noise, is a stochastic signal with constant power spectral density over the entire frequency range. In contrast, colored noise has a non-trivial power spectral distribution, often with a maximum or minimum at certain frequency value.

The performance of a super-resolution microscope can be characterized by spatial resolution. In practice, the achievable resolution depends not only on the optical properties of the system but also on the properties of the fluorophores such as photoswitching and labeling densities  \cite{boden2020predicting}. A straightforward approach to assess the performance or resolution of a STED microscope is to measure the depletion curve, which is defined as the normalized probability that an excited fluorophore contributes to the measured signal as a function of the STED beam intensity  \cite{vicidomini2014importance}. So far, STED microscopy theory considers  the properties of the fluorophore and  its transition under beam illumination to be noise-free. In addition, the potential bias in inhibition of fluorescence by stimulated emission due to fluctuations in the intensity of the STED beam are neglected. To overcome this limitation, we present in this Letter an analytical formulation to measure the depletion function based on colored noise and continuous-wave (CW) depletion beam. 

\subsection{Theory}
\begin{figure}[htbp]
\centering
\includegraphics[width=\linewidth]{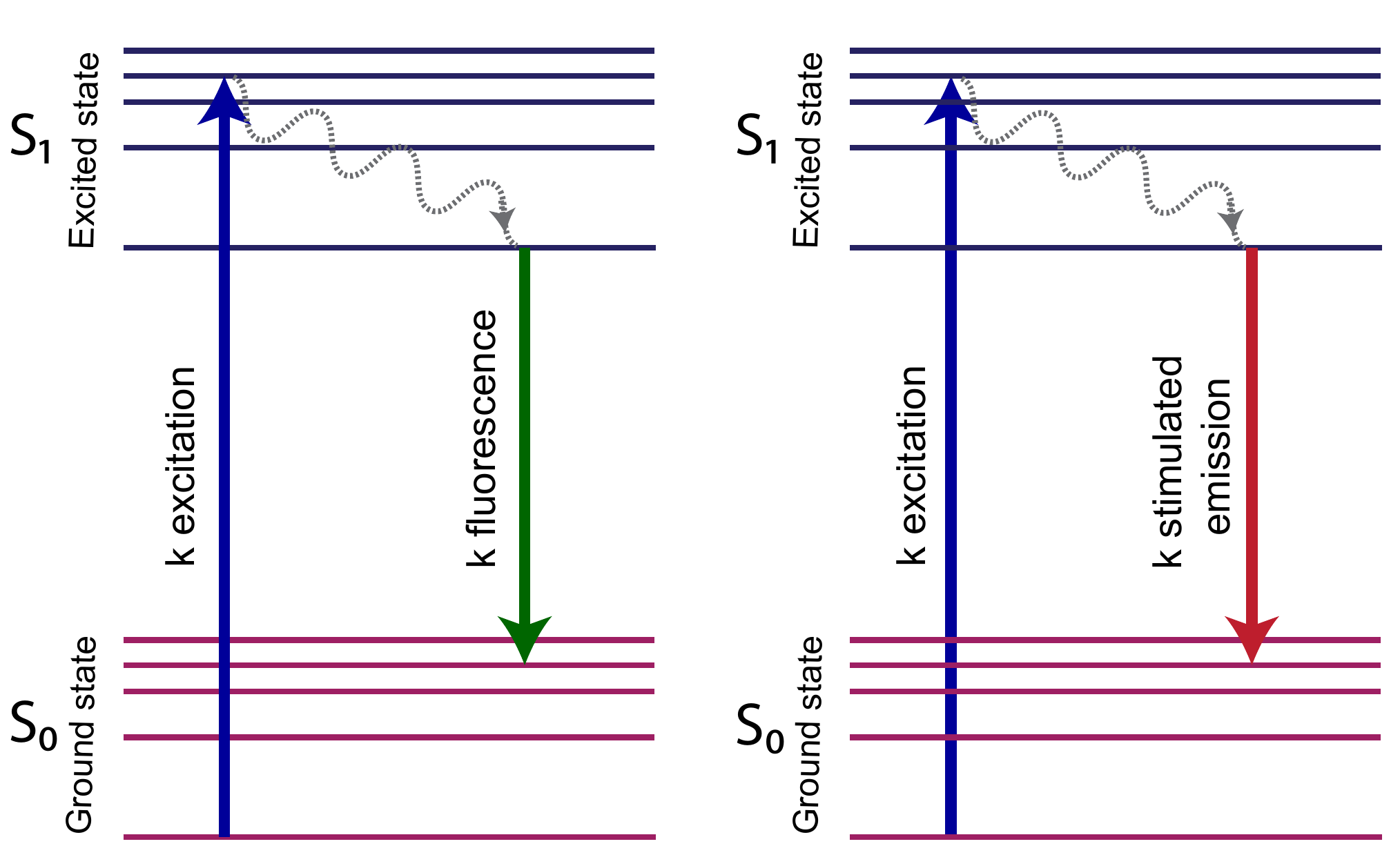}
\caption{Simplified Jablonski energy diagrams showing the processes of excitation, fluorescence and stimulated emission. The fluorophores are excited by an excitation laser. Then, the excited molecules relax to the ground state via spontaneous or stimulated emission induced by a second laser.}
\label{Jablonski}
\end{figure}

In what follows,  we perform the calculation of the depletion, considering that the STED laser has a noisy component. We used a two-step fluorescence mechanism that only takes into account  the ground electronic state $\mathrm{S_{0}}$ and the first electronic excited state $\mathrm{S_{1}}$ of the system (see Fig.~\ref{Jablonski}). 

The fluorescence signal is proportional to $\mathrm{N_{1} (t)}$, which is obtained by solving the following set of stochastic differential equations (SDE) where $\mathrm{N_{0} (t)}$ and $\mathrm{N_{1} (t)}$ are the corresponding ground and excited-state normalized populations, respectively.
\begin{equation}
\begin{split}
\mathrm{ \frac{dN_0(t)}{dt}}=-\mathrm{k_{e} \cdot N_0(t)+k_{s}(t)\cdot N_1(t)+k_{f} \cdot N_1 (t) } \\
\mathrm{ \frac{dN_1(t)}{dt}}=-\mathrm{ k_{f} \cdot N_1(t)-k_{s}(t)\cdot N_1(t)+k_{e} \cdot N_0 (t) } 
\end{split}
\label{eqsys}
\end{equation}
 The parameter $\mathrm{ k_{e}}$ represents the excitation rate from $\mathrm{S_0}$ to $\mathrm{S_1}$ while $\mathrm{k_{f}}$ represents the probability of emitting a photon in the relaxation process from $\mathrm{S_1}$. For simplicity, these two quantities are assumed to be independent of time. The time-dependent $\mathrm{ k_{s}(t)}$ function considers the rate of stimulated emission induced by the depletion laser. Given the linear relationship between $\mathrm{ k_s(t)}$ and the intensity of the depletion laser, $\mathrm{ I_s(t)} $ ($\mathrm{ k_s(t)}=\sigma_\mathrm{s} \mathrm{I_s(t)}$), they will share the same stochastic properties. The proportionality constant $\sigma_\mathrm{s}$ is the cross-section of the stimulated emission. The intensity of the STED laser can be seen as the fluctuation function $\delta\mathrm{I_{s}}(t)$ relative to the average laser intensity $\mathrm{\langle I_{s} \rangle} $, which is time-independent. Note that the latter is also valid for $\mathrm{ k_{s}(t)}$ such that, $\mathrm{ k_{s}(t)} =  \langle \mathrm{ k_{s} } \rangle + \delta \mathrm{k_{s}}(\mathrm{t})$. This, together with the fact that the sum of the populations of $\mathrm{S_{0}}$ and $\mathrm{S_{1}}$ is normalized ($\mathrm{N_0(t)+N_1(t)=1}$), allows for the simplification of the SDE as:   
\begin{equation}
\mathrm{\frac{dN_1(t)}{dt}}= -\left[ \mathrm{ k_{f}+  \left\langle  k_{s} \right\rangle +k_{e} } \right] \cdot \mathrm{ N_1(t) + k_{e} } -\delta \mathrm{k_{s}(t) }  \cdot \mathrm{N_1(t)}
\label{eqs02}
\end{equation}
This equation is a multiplicative noise stochastic differential equation \cite{van1997nonequilibrium,garcia2012noise}, and its solution is the cornerstone of this study. There are mainly two different ways to proceed based on the interpretations of Ito and Stratonovich.  The two are entirely equivalents, the main difference being that the Stratonovich integrals are defined so that the chain rule of ordinary calculus holds. Therefore, we use the Stratonovich interpretation in what follows. 

The \eqref{eqs02} can be rewritten as,
\begin{equation}
\mathrm{ \frac{dN_1(\tilde{t})}{d\tilde{t}}  }= -\left[1+ \frac{\delta \mathrm{ k_{s} (\tilde{t}) } }{ \mathrm{ K } }\right] \mathrm{  N_1(\tilde{t}) } + \mathrm{ \frac{k_{e}}{K} }
\label{eqa}
\end{equation}
where $ \mathrm{ \tilde{t}= K \cdot t }$ is a dimensionless variable and $ \mathrm{ K = k_{f}+\langle k_{s}\rangle+k_{e} } $ represents the sum of all the time-independent rates. From here on, we will omit the tilde mark over "$\mathrm{ t }$" to avoid cumbersome notation.

To proceed, we need to specify the nature of the noise $\delta \mathrm{k_{s}}$, and in our case,  we will consider the well known colored Ornstein-Uhlenbeck noise. This choice enables us to simultaneously study the effects of noise variance and correlation time on the depletion value. The noise $\delta \mathrm{k_{s}}$ can be generated through the Ornstein–Uhlenbeck equation
\begin{equation}
\mathrm{ \dfrac{d}{dt} } \delta \mathrm{ k_{s}(t) } =  -\lambda \cdot \delta \mathrm{ k_{s}(t) \left( t \right) } + \lambda \cdot \beta \mathrm{ \left( t \right) }
\label{eqa18}
\end{equation}
where  $\beta \mathrm{ (t) } $ is a white noise of auto-correlation function $\left<\beta \mathrm{ (t)}\beta \mathrm{(t')}\right>=\Delta \cdot \delta \mathrm{(t-t')}$ and $\Delta$ its variance. The parameter $\lambda = \tau^{-1}$ represents the inverse of the characteristic correlation time. In addition, we simply consider $\delta \mathrm{ k_{s} }(0)=0$ as the initial condition.
This particular model proposed for $\delta\mathrm{k_{s}(t)}$ guarantees that in the limit $t\rightarrow\infty$ and $\lambda\rightarrow\infty$ ( Gaussian white noise limits), $\left<\delta\mathrm{k_{s}(t)}\delta\mathrm{k_{s}(t')}\right>=\Delta \cdot \delta \mathrm{ (t-t')}$. 
It means that by varying the characteristic correlation time in Eq.\ref{eqa18} we can interpolate between the Gaussian white noise limit and a highly time-correlated Ornstein-Uhlenbeck noise. For the last process, a direct calculation \cite{gardiner1985handbook} allows us to obtain the auto-correlation function such as:

\begin{eqnarray}
\nonumber
\left\langle \delta \mathrm{k_{s}} \mathrm{ \left( t \right) } \cdot \delta \mathrm{ k_{s}} \mathrm{ \left( t' \right) }  \right\rangle &=& \frac{\Delta \cdot \lambda}{2}  \mathrm{ exp } \left[   - \lambda \cdot \vert \mathrm{t-t' }   \vert \right] \\ 
&-& \frac{\Delta \cdot \lambda}{2} \mathrm{ exp } \left[   - \lambda \cdot \left(  \mathrm{t+t' }   \right)  \right]
\label{eqa20}
\end{eqnarray}
which in the long time limit ($\mathrm{ t, t'} \gg \lambda^{-1}$) behave as:
\begin{equation}
 \left\langle \delta \mathrm{ k_{s}} \mathrm{ \left( t \right) } \cdot \delta \mathrm{ k_{s}} \mathrm{ \left( t' \right) }  \right\rangle = \frac{\Delta \lambda}{2} \cdot  \mathrm{ exp } \left[- \lambda \cdot \vert \mathrm{t-t' }   \vert \right]. 
\label{eqa21}
\end{equation}

The knowledge of the statistical properties of $\delta  \mathrm{ k_s(t)}$ are crucial for the determination of the average population of our model $\langle  \mathrm{ N_1(t)}\rangle$. This quantity will be used to calculate the depletion, which is defined as:
\begin{equation}
\eta =\frac{\langle \mathrm{N}_1(+\infty,\langle k_s\rangle)\rangle}{\langle  \mathrm{N}_1(+\infty,0)\rangle }.
\label{eqa18a}
\end{equation}
where $\langle \mathrm{N}_1(+\infty,\langle k_s\rangle)\rangle$ corresponds to the average population at $\mathrm{t}\rightarrow\infty$, after taking the statistical  average over realisations of $\delta k_s(t)$ and the normalization factor $\langle  \mathrm{N}_1(+\infty,0)\rangle$ corresponds to the equilibrium population setting $\mathrm{k_{s}=0}$. The long time limit $ \mathrm{t} \rightarrow \infty$ is considered because, experimentally, the depletion is calculated assuming a steady-state condition. Physically $\eta$ give the probability to force fluorophores to their ground off-state for a given STED laser power. 

To calculate the  average value of $\langle  \mathrm{ N_1(t)}\rangle$, we proceed first with the solution of \eqref{eqa}, using standard methods for solving linear differential equations \cite{LevElsgolts}. In a second step, we take the statistical average of the solution, which yields
\begin{eqnarray}
\nonumber
\mathrm{ \left\langle N_{1}(t) \right\rangle } &=& \mathrm{ N_{1} \left( 0 \right) \cdot exp \left(- t \right)  }  \mathrm{I(t,0)} \\ 
&+& \mathrm{ \dfrac{k_{e}}{K} }\int_{0}^{\mathrm{t}}  \mathrm{dt'} \mathrm{exp  \left[ - (t-t')  \right] }  \mathrm{ I(t,t')},
\label{eqa24}
\end{eqnarray}
where $\mathrm{ N_{1} \left( 0 \right)}$ is the initial population and the function $\mathrm{ I(t,t')}$ is given by the general expression
\begin{equation}
\mathrm{ I(t,t') } = \left\langle \mathrm{ exp } \left[   - \int_{\mathrm{t'}}^{\mathrm{t}}  \dfrac{\delta \mathrm{ k_{s} (t_{2}) }}{\mathrm{ K }} \mathrm{ dt_{2}} \right]     \right\rangle 
\end{equation}

To continue with the determination of $\mathrm{ I(t,t')}$, we can now expand the exponential into its power series and
then compute the corresponding average to the $\mathrm{2n}$-point correlation functions of $\delta  \mathrm{k_s(t)}$, evaluated at  different times. Since we assume that $\delta \mathrm{k_s(t)}$ is a Gaussian random variable we can always write the $\mathrm{2n}$-point correlation functions in terms of the two point correlation function $\langle\delta \mathrm{k_s(t_1)}\delta \mathrm{k_s(t_2)}\rangle$. The result obtained then allows a resummation of the series \cite{kardar2007statistical} after the formal integration over the time variables yielding 
\begin{eqnarray}
\mathrm{ I(t,t') } &=&  \mathrm{exp}   \left[ \dfrac{1}{\mathrm{2K^{2}}}  \int_{\mathrm{t'}}^{\mathrm{t}} \int_{\mathrm{t'}}^{\mathrm{t}} \left\langle \delta \mathrm{ k_{s}} \mathrm{ \left( t_{1} \right) }\delta \mathrm{ k_{s}} \mathrm{ \left( t_{2} \right) }  \right\rangle \mathrm{ dt_{1}dt_{2}}  \right].
\label{eqa26}
\end{eqnarray}

If we now take into account the specific form of $\langle \delta \mathrm{ k_{s} } \mathrm{ \left(t\right) } \delta \mathrm{ k_{s}} \mathrm{ \left(t'  \right)} \rangle$ given in \eqref{eqa20}, we can calculate exactly the form of $\mathrm{ I(t,t')}$ and consequently the value of $\langle \mathrm{N_1(t)}\rangle$ in the long time limit ($\mathrm{ t} \rightarrow\infty$). In this limit, we can verify that the contribution from the initial conditions from both $\mathrm{N_1(0)}$ and $\delta\mathrm{k_s(0)}$ are negligible for weak enough noise intensities ($\Delta<\mathrm{ 2K^2}$). The described procedure lead us to
\begin{eqnarray}
\nonumber
\mathrm{ \left\langle N_{1}(+\infty) \right\rangle } &=& \mathrm{ \dfrac{k_{e}}{K} } \cdot \mathrm{exp} \left( -\dfrac{\Delta}{2\lambda \mathrm{K^{2}}}  \right) \int_{0}^{\infty} \mathrm{exp} \left[  - \left( \mathrm{u} - \dfrac{\Delta}{2\mathrm{K^{2}}} \mathrm{u}   \right)  \right]\\
& \cdot &  \mathrm{exp} \left( \dfrac{\Delta}{2\mathrm{K^{2}} \lambda} e^{-\lambda \mathrm{u}}  \right) \mathrm{du}.
\label{eqa30}
\end{eqnarray}
The integral above can be written in terms of Gamma and  incomplete Gamma functions in the form:
\begin{eqnarray}
\nonumber
\mathrm{ \left\langle N_{1}(+\infty) \right\rangle } &=& \mathrm{ \dfrac{k_{e}}{K} } \cdot \mathrm{exp} \left( -\dfrac{\Delta}{2\lambda \mathrm{K^{2}}}  \right) \cdot \left( -\dfrac{\Delta}{2\lambda \mathrm{K^{2}}}  \right)^{\dfrac{-1+\frac{\Delta}{2\mathrm{K}^{2}}}{\lambda}}     \\ \nonumber
&&\dfrac{1}{\lambda} \left[   \Gamma \left(  \dfrac{1-\frac{\Delta}{2\mathrm{K}^{2}}}{\lambda}  \right) -  \Gamma \left(  \dfrac{1-\frac{\Delta}{2\mathrm{K}^{2}}}{\lambda}, - \dfrac{\Delta}{2\mathrm{K}^{2} \lambda} \right)  \right]. \\ 
\label{eqa31}
\end{eqnarray}
Although this expression is not easy to interpret physically, it allows a straightforward calculation of the depletion, as defined in \eqref{eqa18a}. Additionally, we can analyse the limit cases corresponding to $\lambda\rightarrow0$ or $\Delta\rightarrow0$ and $\lambda\rightarrow\infty$. The first scenario corresponds to the ideal case in which noise does not play any role and the second case  when the colored noise becomes a white noise due to a reduction of the correlation time. For those cases, simple analytical expressions for depletion can be obtained allowing a direct physical interpretation of the results.  

\subsubsection{Ideal case, $ \lambda\rightarrow0  $ or $\Delta\rightarrow0$}
In this limit, we recover $\langle\mathrm{N_{1}}(+\infty)\rangle=\mathrm{k_{e}/K}$, which lead us to
\begin{equation}
\eta=\left[ 1+\frac{\langle \mathrm{ k_{s}}\rangle}{\mathrm{ k_{f}}+\mathrm{k_{e}}} \right]^{-1}.
\end{equation}
This result can be expressed in terms of the intensity of saturation $\mathrm{I_{sat}}=\frac{\mathrm{k_{f}}+\mathrm{k_{e}}} {\sigma_s}$, this variable just consider the the fluorescent properties of the molecules. In this way we reach the expression: 
\begin{equation}
\eta=\left[ 1+\frac{\langle \mathrm{ I_{s}} \rangle}{\mathrm{ I_{sat}}} \right]^{-1}.
\label{eqafree}
\end{equation}

The above expression is well know in the super-resolution imaging community and have been used in theoretical and experimental contexts to assess the performance of STED microscopy. 

\subsubsection{White noise case, $\lambda\rightarrow\infty$ }
Using \eqref{eqa31} we can obtain that the average population of the steady-state is given by:
\begin{eqnarray}
\mathrm{ \left\langle N_{1}(+\infty) \right\rangle } &=& \mathrm{ \dfrac{k_{e}}{K} } \cdot \left[ 1-\frac{\Delta}{2\mathrm{K}^{2}} \right]^{-1}
\label{eqa33}
\end{eqnarray}

Although this expression was obtained using, as a premise, the Gaussian character of the coloured noise, it can be shown that in the limit of white noise, such a result holds even if the noise have a non-Gaussian local distribution. We can now proceed with the calculation of the depletion in this scenario, which yields
\begin{equation}
\eta= \left[ 1+\frac{\langle \mathrm{ k_{s} }\rangle-\frac{1}{2}\frac{\Delta}{ \mathrm{ k_{f}+\langle k_{s}\rangle+k_{e}}}}{ \mathrm{k_{f}+k_{e}}} \right]^{-1}.
\end{equation}
This expression shows some limitations of our mathematical model. We can notice that for $\langle \mathrm{ k_{s}}\rangle<\frac{1}{2}\frac{\Delta}{\mathrm{ k_{f}+\langle k_{s}\rangle+k_{e}}}$, the calculated depletion would be greater than one. This is a non-physical result produced by the fact that our mathematical model for $\mathrm{k_s(t)}$ does not rule out the possibility of negative values for this quantity at large enough noise amplitudes once we have fixed $\langle \mathrm{k_{s}}\rangle$. In this nonphysical scenario sufficiently high noise fluctuations in $\mathrm{ k_s(t)} $ can produce the enhancement of the values of $\langle \mathrm{ N }_1(+\infty)\rangle$ when compared to its corresponding value in the absence of the STED laser.

Rewriting the depletion in terms of $\langle \mathrm{ I_{s}}\rangle/\mathrm{I_{sat}}$ we get our working expression for the depletion in the white noise limit:
\begin{equation}
\eta=\left[1+\frac{\langle \mathrm{ I_{s}}\rangle}{\mathrm{I_{sat}}}-\frac{\alpha^2}{2}\frac{\left(\frac{\langle \mathrm{ I_{s}}\rangle}{\mathrm{I_{sat}}}\right)^2}{1+\frac{\langle \mathrm{ I_{s}}\rangle}{\mathrm{I_{sat}}}}\right]^{-1},
\label{dep2}
\end{equation}
where $\alpha=\frac{\sqrt{\Delta}} {\langle \mathrm{ k_{s}}\rangle}$, is a quantity that characterizes the laser noise distribution giving us the relative standard deviation (rsd) of the STED laser. This equation links the rsd of the laser to the final depletion efficiency obtained at a given STED power.
This model is expected to better describe the CW-STED depletion curve as it depends not only on the sample properties but also on the noise properties of the depletion laser. In the absence of laser noise $(\Delta=0)$, Eq.~\ref{dep2} reduces to the well-known noise-free depletion expression for CW-STED microscopy.

 \begin{figure}[ht]
\centering
\includegraphics[scale=0.4]{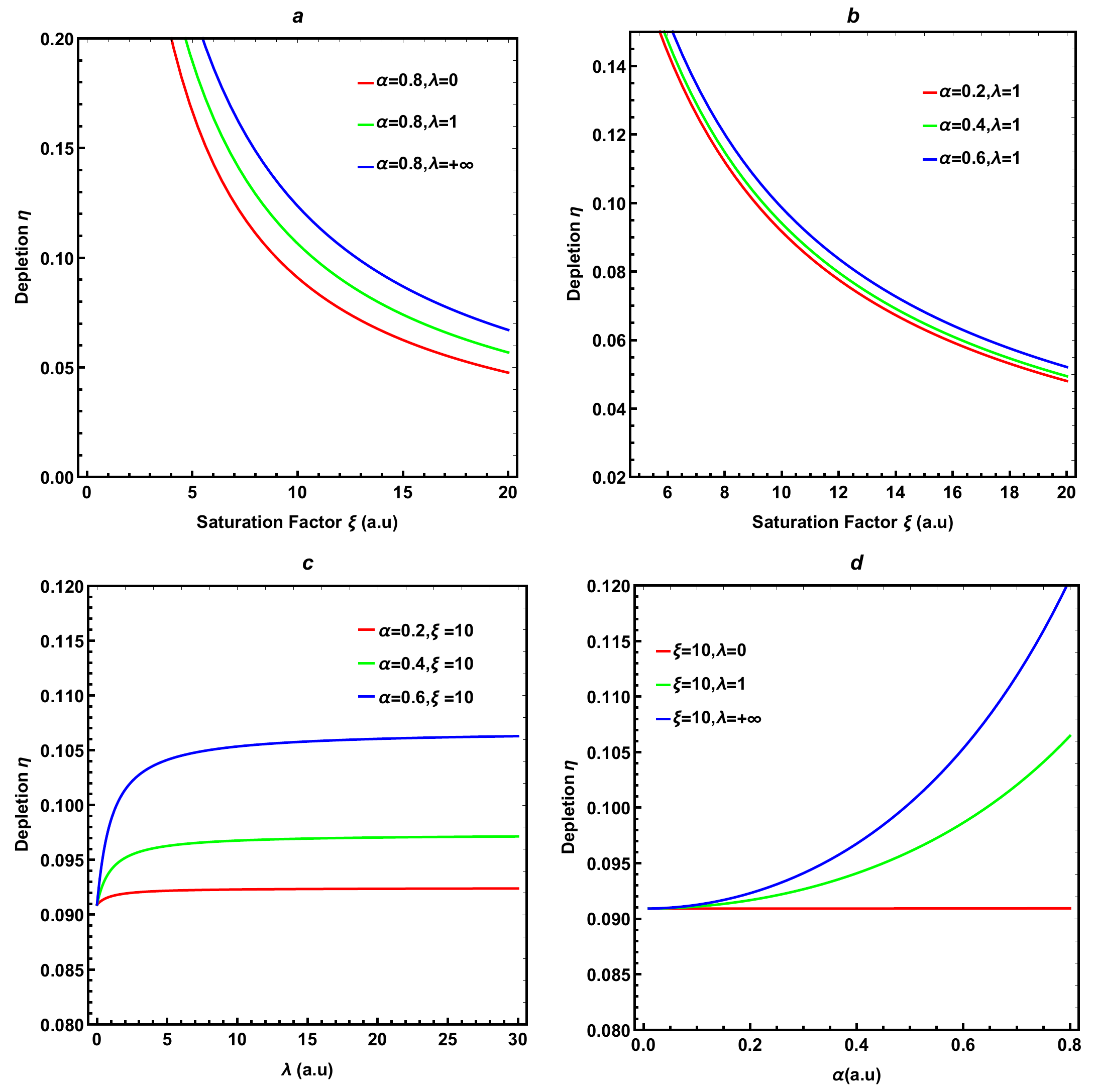}\caption{Theoretical calculation of the laser intensity noise effect on the performance of the CW-STED microscopy. (a) Depletion curves corresponding to coloured noises with different correlation times, the corresponding values of the characteristic correlation time are indicated in the inset of the figure. (b) Depletion curves for different noise strengths (0.2 rsd, 0.4 rsd and 0.6 rsd) fixing the correlation time of the coloured noise. (c) and (d) Behavior of the depletion efficiency for a given saturation factor ($\xi$=10) varying the noise strength and the correlation time, respectively.}
\label{Figure 2}
\end{figure}
 
In Fig. \ref{Figure 2}, we study the behaviour of the depletion varying the strength of the noise ($\Delta$) and  the inverse of the characteristic correlation time ($\lambda$) in different scenarios. The depletion as a function of the saturation factor $\frac{\langle \mathrm{I_{s}}\rangle} {\mathrm{I_{sat}}}$, represented as $\xi$ in what follows was studied numerically. The fig.~\ref{Figure 2}~(a) shows the behaviour of the depletion ($\eta$) in three different scenarios: ideal case ($\alpha=0$), coloured noise ($\alpha\neq0$ and $0<\lambda<+\infty$) and white noise ($\lambda\rightarrow\infty$). As expected, the optimal depletion curve corresponds to the ideal case, in which the laser noise is absent~\cite{coto2014influence}. On the other hand, when the noise is present our theory confirms that the higher the intensity noise the lower the depletion efficiency for a given average intensity. For instance, a noise intensity of $0.6$ rsd will need a saturation factor of roughly $20$ to reach the same fluorescence quenching obtained with a saturation factor of $14$, in the case of a noise intensity of $0.2$ rsd (figure 2(a)). The presence of fluctuations deteriorates the depletion i.e. the higher the laser stability, the higher the depletion efficiency, see Fig.~\ref{Figure 2}~(b). In the low-intensity noise regime $(\mathrm{rsd<0.2})$, depletion efficiency is not significantly affected. However, when the strength of the noise is high enough ($\mathrm{rsd>0.6}$), the decrease in efficiency can no longer be neglected. An increase of the variance of the noise results in a suboptimal depletion efficiency and an increase of the intensity of the STED beam is needed to recover the depletion efficiency of low noise scenario. In Fig.~\ref{Figure 2} (c) and (d), we observe that at a given saturation factor the depletion is strongly affected by an increase of $\alpha$ and the inverse of the correlation time of the noise ($\lambda$), in a way that systems with a higher $\alpha$ ($\lambda$) are more affected by an increase of $\lambda$ ($\alpha$).  

\subsubsection{Simulation}
\begin{figure}[ht]
\centering
\includegraphics[width=\linewidth]{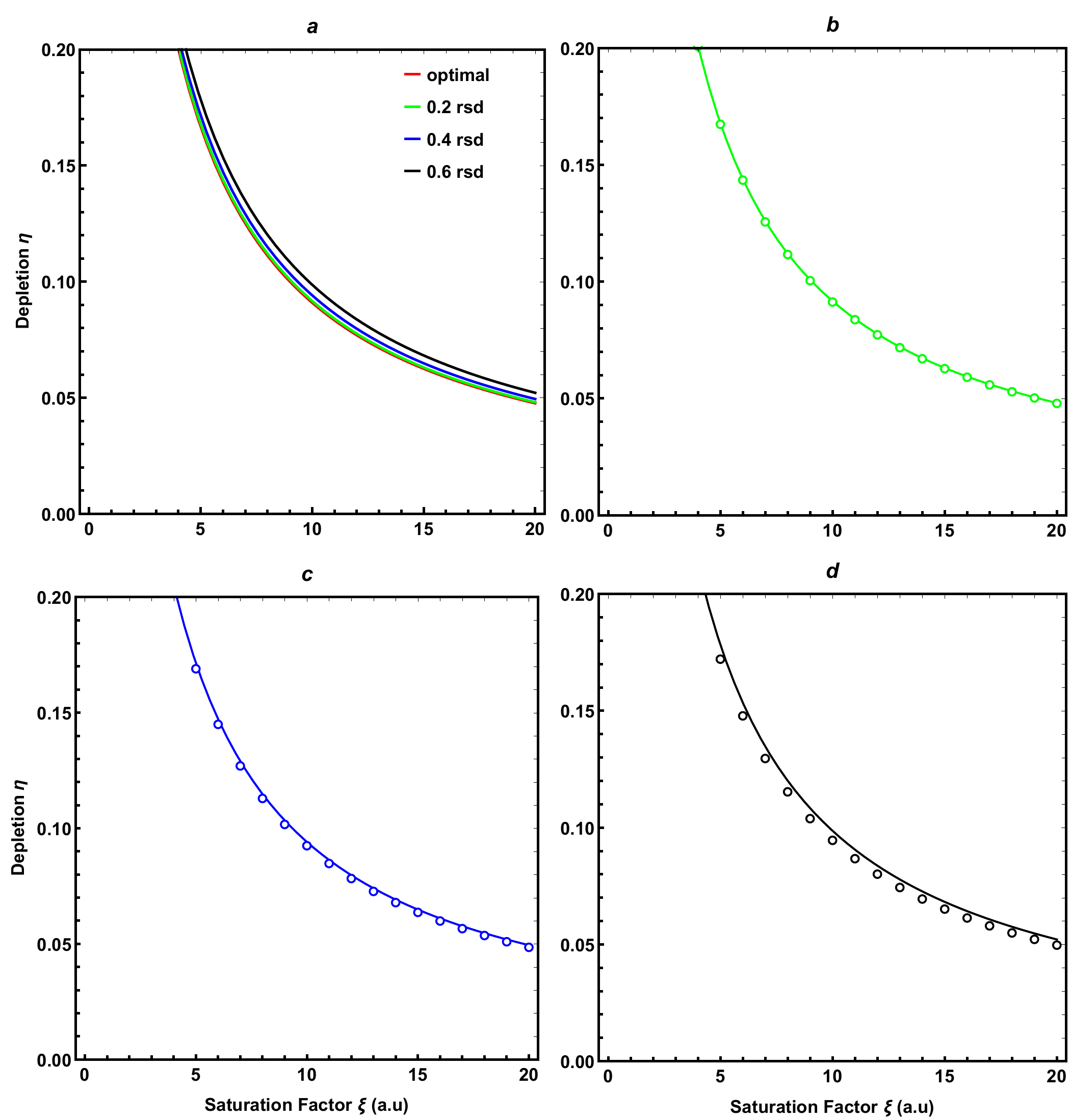}
\caption{Influence of colored noise intensity on depletion efficiency of CW-STED microscopy. (a) Theoretical depletion curves at different noise strengths (0.2 rsd, 0.4 rsd, and 0.6 rsd). (b-d) Comparison between analytical (full line) and computational (empty circles) simulation results for depletion as a function of saturation factor. The relative difference between theoretical and computational results for the different noise strengths 0.2 rsd, 0.4 rsd and 0.6 rsd was less than $2\%$, $3\%$ and $5\%$, respectively.}
\label{Figure 3}
\end{figure}

The analytical predictions obtained until here were verified through the numerical solution of Eq.~\ref{eqa} using well-established routines developed in the Mathematica 11 software \cite{Mathematica}.  The stationary population average and consequently the numerical depletion was estimated considering a large number of noise realizations $\beta\mathrm{(t)}$. The latter was taken so that the difference between the analytical prediction and the corresponding numerical average is always less than $5\%$. Fig.~\ref{Figure 3} b-d shows the comparison between the analytical and numerical results for the general case of colored noise at different noise strengths. As can be seen, a good agreement is obtained, which validates the analytical predictions. In addition, we perform simulations for the limit case of white noise (results not shown), obtaining the same level of agreement.

\subsubsection{Comparison of theory with experimental results}

\begin{figure}[htbp]
\centering
\includegraphics[width=\linewidth]{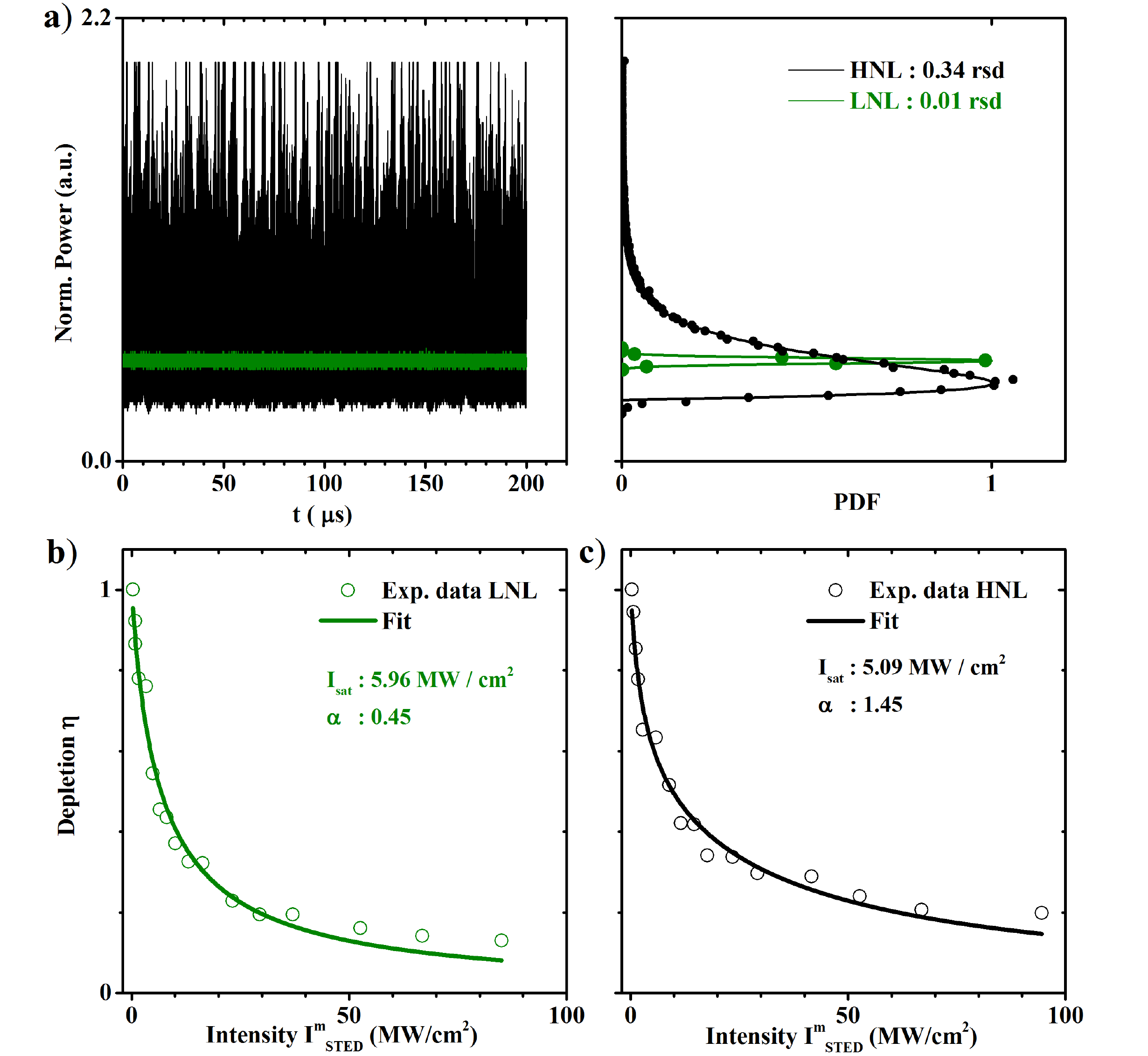}
\caption{Effect of laser fluctuations in CW-STED microscopy. (a) Characterization of the intensity of the two lasers based on representative time traces of 200 µs length (sampling 1 ns). The right panel displays the normalized Probability Density Functions of the two lasers. (b-c) Depletion curves measured with Alexa 488-labeled antibody (empty dots) and their corresponding fit (lines) according to the white-noise model proposed in Eq. \ref{dep2}.}
\label{expdata}
\end{figure}

Having validated our analytical model, we fit experimental depletion curves, previously published \cite{coto2014influence}, to the analytical expressions. For simplicity, we used only those obtained for the white noise limit, Eq.~\ref{dep2}. Two CW depletion lasers with the same average intensity but with different intensity noise profiles are investigated,  see Fig.~\ref{expdata}. The Low Noise Laser (LNL) has a normal distribution (0.01 rsd), while the High Noise Laser (HNL) has an unknown distribution (0.34 rsd). The measurement time was long enough to assume  that the equilibrium was reached. From the fit of the experimental curves, we extracted the saturation intensity ($\mathrm{I_{sat}}$) of the fluorophore and the noise (rsd) of the STED laser. No significant changes were found for low-noise scenarios when fitting the experimental curve with our model and the noise-free depletion curve, Eq.~\ref{eqafree}. These two models are statistically consistent.  Previous work~\cite{coto2014influence} empirically introduced a constant offset $\alpha$ in the noise-free depletion curve model, i.e., $\eta_{\mathrm{noise}}(\mathrm{I_{s}})=(1-\alpha)\eta(\mathrm{I_{s}})+\alpha$ to explain reduction of depletion efficiency for high-noise scenarios.  It should be noted that the model obtained here offers an analytical expression for such scenarios. The experimental depletion curves are well described by the model with fitted parameters $I_{\mathrm{sat}} = 5.09 \mathrm{MW cm^{-2}}$ and $\alpha= 1.45$ for the HNL and $I_{\mathrm{sat}} = 5.96 \mathrm{MW cm^{-2}}$ and $\alpha= 0.45$ for the LNL. On the other hand, at high intensities, there is a small deviation between our theoretical curves and those obtained experimentally. A possible explanation for this effect is related to the incomplete decay of the depletion curves due to the signal background caused by the excitation from the STED beam (anti-Stokes emission) \cite{coto2014new}. Overall, the model works well for both scenarios, high and low noise, with an adjusted R-squared value of 0.98, giving a good agreement with the previously published experimental data \cite{coto2014influence}. Finally, since the colored noise model has the Gaussian white noise model as a limiting case, it is expected that the fits using the colored noise will also work. At least we will always have the trivial solution in which the correlation time resulting from the fit is a small quantity. On the other hand, given that the experimental data have a non-negligible noise level, it will be impossible to establish which of the two theoretical models is more appropriate to describe the experiments. The chi-square test in both cases yields similar values. 

 \subsubsection{Discussion and Conclusions}
 
This letter theoretically demonstrated the importance of using stable lasers to reduce the sample illumination on CW-STED implementations. The use of a noise-eater is strongly recommended to stabilize the amplitude of a high-noise depletion laser. On the other hand, laser power stability lower the intensity to reach a certain resolution \cite{Leutenegger2010, Moffitt2011, Vicidomini2013}, thus they reduce potential photodamage effects and re-excitation caused by the depletion laser \cite{coto2014new}. 

As we have shown, intensity fluctuations play a negative role in the performance of CW-STED microscopy. However, controlled variations of the STED intensity induces spatially encoded variations of the fluorescence emission that can, in principle, be decoded to further improve the effective spatial resolution of the STED image \cite{sarmento2018exploiting, lanzano2015encoding}. As a result, if these fluctuations are adequately detected, one can exploit the 'natural' changes of STED intensity during the image acquisition and separate photons based on the depletion dynamics in the phasor plot.
 
In conclusion, this work introduces an analytical formulation capable of accurately describe the impact of intensity fluctuations and intensity correlation time on the performance (depletion efficiency) of a CW-STED microscope. The effects of noise intensity on image resolution can be understood by consdering the linear proportionality relation ship of this quantity with the depletion efficiency \cite{vicidomini2014importance}. Comparison with numerical simulations and previously published experimental data validated the analytical results. The analytical approach followed here can easily be extended to other imaging modalities, such as ground-state depletion and RESOLFT microscopy \cite{hell1995ground, testa2012nanoscopy}. In future works, we will investigate the effects of time jitter and donut variability (shape and polarization) on efficiency of the STED microscope \cite{ neupane2013tuning}.\\ 
\textbf{Acknowledgment.}
The Berthiaume Family Foundation supported this study. In addition, the authors thank Giuseppe Vicidomini (Istituto Italiano di Tecnologia) and Luca Lanzano ( University of Catania) for helpful comments on the article. We also thank Nate Jowett for proofreading the manuscript. Finally, A.M.C. acknowledges financial support from Funda\c{c}\~ao de Amparo \`a Pesquisa de Santa Catarina FAPESC.  \\
\textbf{Disclosures.} The authors declare no conflicts of interest.
\\
\textbf{Data Availability.} Data underlying the results presented in this paper are not publicly available at this time but may be obtained from the authors upon reasonable request.
\bibliography{Bibliography}

\end{document}